\documentclass[oneside,11pt]{article}

\usepackage[utf8]{inputenc}
\usepackage[T1]{fontenc}

\usepackage{amsthm}
\usepackage{amsmath}
\usepackage{amsfonts}
\usepackage{amssymb}

\usepackage{graphicx} 
\usepackage{float}
\usepackage{booktabs}
\usepackage{multirow}
\usepackage{rotating}
\usepackage{array}
\usepackage{xcolor}
\usepackage{animate}

\usepackage{breakcites}

\usepackage{enumitem}

\usepackage[top=1.5cm,bottom=2cm,right=2.5cm,left=2.5cm]{geometry}
\usepackage{titling}
\usepackage{natbib}

\usepackage{ifplatform}

\usepackage{textcomp}

\ifwindows
\fi

\ifwindows
	\usepackage{bbm}
\else
	\iflinux
		\usepackage{bbm}
	\else
		\usepackage{bbm}
	\fi
\fi

\DeclareFontFamily{OT1}{pzc}{}
\DeclareFontShape{OT1}{pzc}{m}{it}{<-> s * [1.10] pzcmi7t}{}
\DeclareMathAlphabet{\mathpzc}{OT1}{pzc}{m}{it}

\usepackage[pdftex,final,bookmarksnumbered,bookmarksopen=false,breaklinks,colorlinks]{hyperref}
\hypersetup{
	final,
    bookmarksnumbered=true,
    bookmarksopen=true,
    bookmarksopenlevel=0,
    unicode=false,          
    pdftoolbar=true,        
    pdfmenubar=true,        
    pdffitwindow=false,     
    pdftitle={Distance weighted discrimination of face images for gender classification},    
	pdfdisplaydoctitle=true, 
	pdftoolbar=true,
	pdfmenubar=true,
	pdflang={English},
    pdfauthor={Monica Benito, Eduardo Garcia-Portugues, J. S. Marron, Daniel Pena},     
    pdfsubject={arXiv paper},   
    pdfcreator={Eduardo Garcia-Portugues},   
    pdfproducer={Eduardo Garcia-Portugues}, 
    pdfkeywords={Distance weighted discrimination}{Feature extraction}{Fisher's linear discriminant}{Gender classification}{HDLSS}{Support vector machines}, 
    pdfnewwindow=true,      
    breaklinks=true,
    hidelinks,       
    linkcolor=black,          
    citecolor=black,        
    filecolor=magenta,      
    urlcolor=cyan           
}

\allowdisplaybreaks

\newif\ifmain
\maintrue
\ifmain

\else

\fi
\newif\ifsupplement
\supplementtrue

\begin{document}

\ifmain

\title{Distance weighted discrimination of face images\\ for gender classification}
\setlength{\droptitle}{-1cm}
\predate{}%
\postdate{}%

\date{}

\author{M\'onica Benito$^1$, Eduardo Garc\'ia-Portugu\'es$^{1,4}$, J. S. Marron$^2$, and Daniel Pe\~{n}a$^{1,3}$}

\footnotetext[1]{
Department of Statistics, Carlos III University of Madrid (Spain).}
\footnotetext[2]{
Department of Statistics and Operations Research, University of North Carolina at Chapel Hill (USA).}
\footnotetext[3]{
Institute of Financial Big Data, Carlos III University of Madrid (Spain).}
\footnotetext[4]{Corresponding author. e-mail: \href{mailto:edgarcia@est-econ.uc3m.es}{edgarcia@est-econ.uc3m.es}.}

\maketitle


\begin{abstract}
We illustrate the advantages of distance weighted discrimination for classification and feature extraction in a High Dimension Low Sample Size (HDLSS) situation. The HDLSS context is a gender classification problem of face images in which the dimension of the data is several orders of magnitude larger than the sample size. We compare distance weighted discrimination with Fisher's linear discriminant, support vector machines, and principal component analysis by exploring their classification interpretation through insightful \textit{visuanimations} and by examining the classifiers' discriminant errors. This analysis enables us to make new contributions to the understanding of the drivers of human discrimination between males and females.
\end{abstract}

\begin{flushleft}
\end{flushleft}

\section{Introduction}

Image classification is one of the most important applications of image analysis and, especially during the last two decades, has received significant attention. In particular, automatic identification of human faces from a database of digital images has become increasingly important. Surveys in face recognition can be found in \cite{Samal1992}, \cite{Valentin1994}, and \cite{Zhao2003}, whereas \cite{Kawulok2016} presents a compendium of recent research trends in the area. The process of facial recognition has been thoroughly investigated by cognitive psychologists \citep{Bruce1988} and questions such as how the brain recognizes a face as male or female have been addressed \citep{Burton1993}, although the mechanisms involved in processing the visual information remain unclear \citep{Wu1990}.\\

In the face classification problem, a certain number of variables are measured in a sample of individuals and the goal is to build a rule to classify a new face within given groups, for example males and females. The first approach for face classification, due to \cite{Galton1910}, was to use certain characteristics of the population, such as angles and distances between facial landmarks. Later, other approaches based on anatomical considerations were developed for the classification of facial features \citep{Bruce1992, Farkas1987, Gizatdinova2010}. However, combining these into a single dataset is highly problematic: different landmark points are used and distinct measurements are made. A third approach has been based on the whole facial image, instead of upon such descriptive measurements. This approach is facilitated by a suitable database with similar viewpoint, pose, and illumination conditions, for effective classification. An example of such a database is shown in Movie \ref{mov:faces}.\\

A typical first step in using full images for face recognition is to \emph{rasterize} each image into a vector by, say, column concatenation, and set each entry of the vector as the gray level at a pixel. Let $X$ be the matrix whose columns are these image vectors. Thus $X$ is $d\times n$, where $n$ is the sample size and $d$ is the dimension of the image vectors, \textit{i.e.} the number of pixels. The task of developing a rule to distinguish males from females is a binary classification problem that is often tackled by \emph{linear classifiers}, \textit{i.e.}, classifiers which find a hyperplane that separates the two classes by partitioning the data space. This is the case for all the classifiers considered in this paper. \cite{Turk1991} used Principal Component Analysis (PCA) to obtain a reduced rank representation of the faces and used the minimum distance classification (\textit{i.e.}, the first nearest neighbor rule) in that eigenspace, to classify males and females. \cite{Belhumeur1997} applied Fisher's Linear Discriminant (FLD) function to this problem. They showed that FLD outperformed PCA, perhaps not surprisingly since the latter does not use class information at all. However, as the problem of face classification is a High Dimension and Low Sample Size (HDLSS) context, FLD encounters two problems. First, the traditional algorithm cannot be used directly because the estimated within-class scatter matrix is always singular. Second, the high-dimensional image vectors can lead to computational challenges. In order to avoid these difficulties, \cite{Yang2003} first used PCA for dimensionality reduction, and then applied FLD, hence using class information in the second step.\\

Support Vector Machines (SVM), originally developed by \cite{Vapnik1995}, are linear classifiers that belong to the class of maximum margin classifiers, \textit{i.e.}, the ones which maximize the minimum distance between the two convex hulls of the data points from each class \citep{Hastie2009}. \cite{Marron2007} showed that for HDLSS data SVM has the problem that a large portion of the data are support vectors, \textit{i.e.}, data points which lie on the margin boundaries. Thus, when the observations are projected onto the normal vector of the hyperplane, many of the projections are identical -- a property called \emph{data piling} -- resulting in reduced generalizability of the classifier. \cite{Ahn2010} showed the existence of directions, where data from one class project to a single point, and from the other class project to another single point. Among such directions, the one with greatest distance between these points is called the \emph{Maximal Data Piling} (MDP) direction. An important precursor of the MDP direction is the \emph{Minimum Projected Kurtosis Direction} of \cite{Pena2000,Pena2001}. The formula for MDP is quite similar to that for FLD, with the pooled within class covariance matrix replaced by the overall covariance matrix. Some algebra reveals that indeed in non-HDLSS situations, the MDP and FLD directions are the same, despite their apparently different algebraic form and quite different behavior in HDLSS contexts. The data piling tendency of SVM in HDLSS contexts motivated \cite{Marron2007} to develop Distance Weighted Discrimination (DWD), which was shown, see also \cite{Hall2005}, to be better than both SVM and FLD for HDLSS data. See \cite{Marron2015} for an accessible review on DWD, and see \cite{Carmichael2017} for a deep analysis of the relationship between SVM and MDP.\\

The main contribution of this paper is twofold. First, we do a head-to-head comparison between DWD and the more commonly used SVM by the use of insightful \emph{visuanimations} \citep{Genton2015} on the discriminant directions (Section \ref{sec:exp:1}), and by the examination of the classical discrimination errors on a separate testing dataset (Section \ref{sec:exp:3}). We complement this comparison with PCA (for the discriminant directions) and FLD (for the discrimination error rate) in order to show the better performance of DWD and SVM. Second, we present in Section \ref{sec:exp:2} some novel approaches to deeply understand the canonical differences between male and female faces (which seems automatic by the human perceptual system) that stem from a careful analysis of the DWD classification rule.

\section{Experimental results}
\label{sec:exp}

In order to demonstrate the effectiveness of DWD for facial pattern recognition, we ran a series of experiments and compared our results with those obtained using PCA, FLD and SVM. The experiments were carried out using a training dataset of frontal view face images of former students from Carlos III University of Madrid. The images were recorded with a digital camera, with somewhat similar illumination conditions. The dataset is comprised of $n=108$ face images of size $IJ$, where $I=248$ and $J=186$. The number of men and women are $n_{1}=54$ and $n_{2}=54$, respectively. Rasterization of the images yields vectors in the $\mathbb{R}^{d}$ space, $d=IJ=46148$, whose entries are the gray level values of all the pixels. To eliminate spurious effects due to the location of each face in the image, we registered them using landmarks. Three landmarks, representing the eyes and nose, were automatically selected, and optimal translations and rotations were then computed using a generalized Procrustes analysis, as described in \cite{Benito2005}. Movie \ref{mov:faces} shows the results of this registration process. Note that in the raw images, the faces move around quite a bit, while they are much stable in the registered image, and thus are more suitable for addressing the classification challenge.

\renewcommand{\figurename}{Movie}
\setcounter{figure}{0}
\begin{figure}[h]
\centering
\animategraphics[autopause,loop,controls,width=0.8\textwidth,buttonsize=2em,trim=0 1cm 0 1.5cm]{3}{images/image_}{1}{108}
\caption{\small Raw and registered images. The registered images were obtained by a generalized Procrustes analysis based on three landmarks: eyes (green) and nose (red). It shows that the registered images are shape-comparable and ready for posterior statistical treatment. A blurring has been applied to preserve anonymity. Movie file available as Supporting Information.\label{mov:faces}}
\end{figure}
\renewcommand{\figurename}{Figure}
\setcounter{figure}{1}

\subsection{Face classification by DWD, PCA, and SVM}
\label{sec:exp:1}

We first ran an experiment in which the discriminant vectors for DWD, PCA, and SVM are computed to separate the training dataset into two groups, men and women. Visualization of these directions gives valuable insights on the discrimination ability of each classifier. Movie \ref{mov:projns} gives a graphical summary of the classification performance of DWD (\hyperref[mov:projns:DWD]{2a}), PCA (\hyperref[mov:projns:PCA]{2b}), and SVM (\hyperref[mov:projns:SVM]{2c}) by means of a comparable march along the image projections in the three discrimination directions (green vertical bar, right panels) and the corresponding projected images (left panels). The horizontal span of the right plots was set to 125\% of the range of the data for the sake of a better visualization.\\

The right panel of Movie \hyperref[mov:projns:DWD]{2a} shows the projections of each data vector onto the DWD direction, where the females are red plus signs, and the males are blue circles. In this and the next plots, the heights of the symbols reflect order in the dataset (for visual separation) and the curves are kernel density estimates, with black showing the overall density, and the red and blue showing the corresponding, group-size adjusted, subdensities for the two subpopulations. Females clearly lie to the left, and males to the right. Near the middle, the faces appear quite androgynous, depicting a continuum separation of genders. Note that none of the faces in the left panel are members of the dataset, but instead are reconstructions of points in the image space that lie along the DWD direction. DWD was implemented using $C=100$ as the penalty parameter based on the suggestions in \cite{Marron2007}.\\

\renewcommand{\figurename}{Movie}
\setcounter{figure}{1}
\begin{figure}[!htpb]
\centering
\begin{animateinline}[autopause,loop,controls,every=2,buttonsize=2em]{2}
  \multiframe{101}{i = 1 + 1}{%
  \vbox{%
	\begin{minipage}[c]{0.1\textwidth}
	\textbf{(2a) DWD} 
	\end{minipage}
	\begin{minipage}[c]{0.8\textwidth}
	\includegraphics[width=\textwidth, trim ={0 1cm 0 1.25cm}, clip]{images/DWD_projection_\i}
	\end{minipage}\\
	\begin{minipage}[c]{0.1\textwidth}
	\textbf{(2b) PCA} 
	\end{minipage}
	\begin{minipage}[c]{0.8\textwidth}
	\includegraphics[width=\textwidth, trim ={0 1cm 0 1.25cm}, clip]{images/PCA_projection_\i}
	\end{minipage}\\
	\begin{minipage}[c]{0.1\textwidth}
	\textbf{(2c) SVM} 
	\end{minipage}
	\begin{minipage}[c]{0.8\textwidth}
	\includegraphics[width=\textwidth, trim ={0 1cm 0 1.25cm}, clip]{images/SVM_projection_\i}
	\end{minipage}
  }}
\end{animateinline}
\caption{\small Projection of the data image vectors onto the DWD, PC1, and SVM discriminant directions, showing the continuous separation between females (red plusses) and males (blue circles). Green line indicates a march along the projections in the discriminant direction, with corresponding reconstructed images on the left. DWD shows a good separation of females and males, going through androgynous. PC1 has some discrimination capability, but not as strong as DWD. SVM shows good male-female separation although some facial features are more clear in the DWD plot, and SVM overfits in terms of data piling. Movie files and a static comparison figure available as Supporting Information. \label{mov:projns}\label{mov:projns:PCA}\label{mov:projns:DWD}\label{mov:projns:SVM}}
\end{figure}
\renewcommand{\figurename}{Figure}
\setcounter{figure}{2}

Movie \hyperref[mov:projns:PCA]{2b} shows the projections of the image vectors onto the PC1 (first principal component) direction. While there is some grouping of males to the right, and females to the left, there is also quite a lot of overlap, especially compared to Movie \hyperref[mov:projns:DWD]{2a}. The inefficiency of PCA at separation of males and females is also clear from the sequence of faces in the left panel, which are less distinctly recognizable as males and females. This is not surprising: PCA does not make use of class information at all, but instead targets only maximal variation in the data. This is apparent from the span of the projections, from $-50$ to $40$ for PCA, versus $-17$ to $12$ for DWD. A close look at the sequence of images shows a strong difference in the lower background part for PCA -- related with the absence/presence of long hair -- which is much more constant in the DWD direction. In fact, this shows that the male-female difference is an important component of the single largest mode of variation in the data.\\

The projections for SVM are shown in Movie \hyperref[mov:projns:SVM]{2c}. This gives a much better separation of the classes than PCA. However, care is needed here, because it also shows the problem of data piling discussed in the introduction. In particular, many red points are piled at the right end of the red distribution, and blue points are piled at the left end of the blue distribution. \cite{Marron2007} showed that this results in overfitting by the SVM. In contrast, Movie \hyperref[mov:projns:DWD]{2a} shows a more efficient discrimination of the two groups and no data piling, which is the result of each data point playing a role in finding the discriminant direction in the data. An additional appealing feature of the DWD projected class distributions is their Gaussian shape, a critical property in the good performance of DWD in applications such as microarray batch adjustment \citep{Benito2004}. It is also interesting to compare both methods in terms of the gap between the modes of the subdensities. The gap for SVM in Movie \hyperref[mov:projns:SVM]{2c} goes, roughly, from $-3$ to $3$, while the gap for DWD is $-5$ to $7$. Bigger gaps are better here, since they indicate real population differences, instead of spurious sampling artifacts driven by overfitting. SVM was fit using $C=1000$ following \cite{Gunn1998}'s recommendation.\\

Finally, note that although the SVM faces in the left panel are recognizable as female in the left and male in the right, the DWD images in Movie \hyperref[mov:projns:DWD]{2a} are sharper and give an overall impression of a better male-female separation. In particular, in DWD faces the outline of the faces is more distinct, the forehead seems more rounded for women and squared for men, the eyebrows are hairier in men, the women's eyes are more open, and women smile a little more. All of these ideas suggest that when a new face image is classified the performance of DWD will be superior to the SVM classification. This will in fact be seen in Section \ref{sec:exp:3}.

\subsection{Face recognition insights from DWD}
\label{sec:exp:2}

Figure \ref{fig:LoadingsPlots} gives insights into face recognition through a new image aimed to highlight the facial aspects that drive gender classification. The entries of the DWD vector (\textit{i.e.}, loadings) are coded with colors, white for $0$, darker shades of red for stronger negative (associated to females) and darker shades of blue for stronger positive (associated to males). These contrasts on the DWD vector show that the more discriminant facial aspects are the eyebrows, eyes, nose, lips, chin, and the outline of the head (in which the presence of short/long hair plays a key role).\\

We focus next on these facial aspects by repeating the DWD analysis for the same dataset, but this time restricting the input vectors to the most relevant features. In particular, we use only pixels that lie within rectangles chosen as in \cite{Benito2005}, which correspond to the eyebrows, eyes, nose, lips, and chin. Figure \ref{fig:ScoresMatrix} shows the result of projecting the full image vectors into each of the DWD directions, that were independently obtained for each of these face regions. The first direction is computed just using the eyebrows, the second one is obtained by only considering the eyes, and the third, fourth and fifth are those corresponding to the nose, lips, and chin, respectively.\\

\begin{figure}[htpb!]
\centering
\includegraphics[width=0.7\textwidth, trim = {0 1cm 0 1cm}, clip]{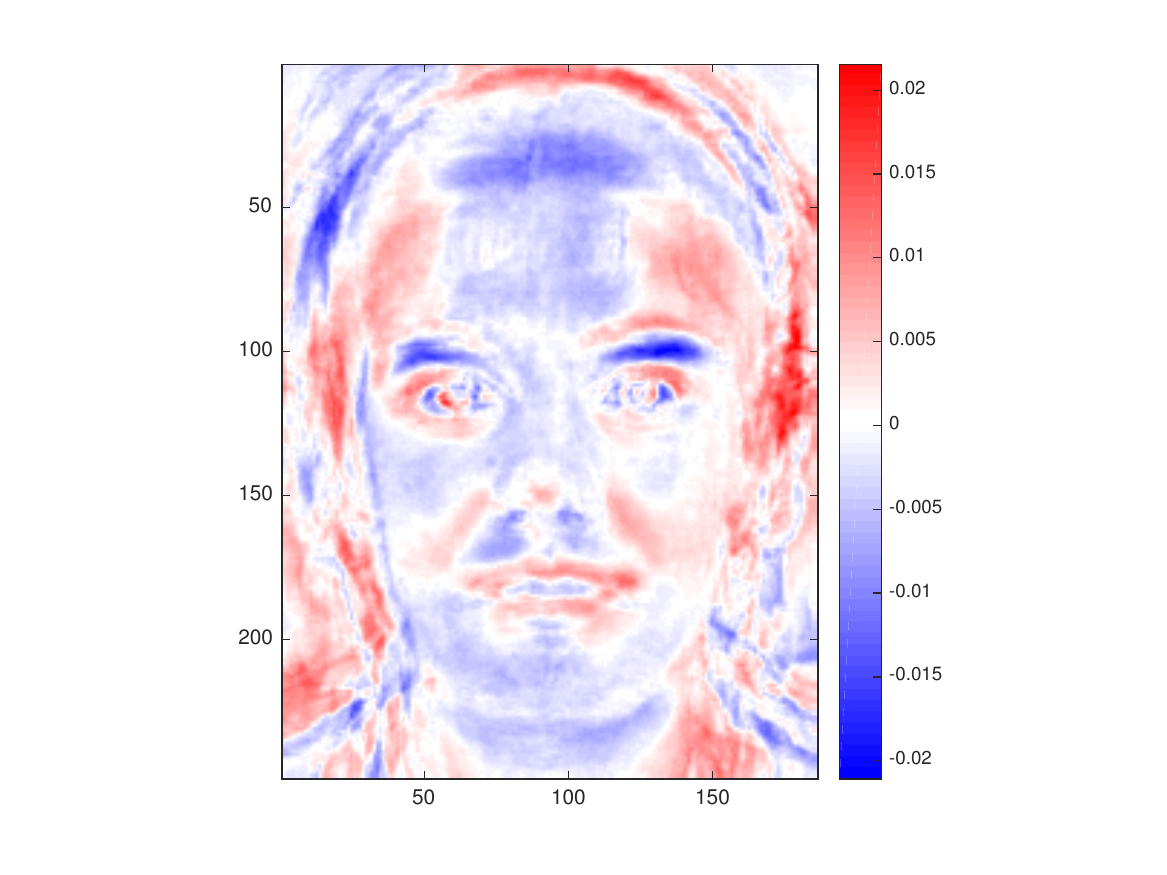}
\caption{\small Highlighting of discriminant pixels in the gender classification problem using DWD loadings. White codes no difference, shade of red (blue) shows magnitude of negative (positive) entry in the DWD vector, associated to females' (males') features. It shows that head shape, together with pixels in the eyebrows, eyes, nose, lips, and chin are important to the discrimination. \label{fig:LoadingsPlots}}
\end{figure}

The univariate distributions shown in the diagonal plots of Figure \ref{fig:ScoresMatrix} indicate how well each region classifies males vs. females using only pixels in that region. Note that both the eyebrows' and eyes' pixels alone produce a complete separation of the subpopulations. For the other regions, we observe some small overlap. The off-diagonal plots are the corresponding pairwise scatterplots: in the second column of the first row, the two-dimensional projection is of the full image vectors on the DWD directions found by using the eyebrows and eyes, respectively. Note that there is even better discrimination in the two dimensional plot, than in either one dimensional version. Also insightful are the black lines indicating the two DWD directions, showing that they are nearly orthogonal, \textit{i.e.} that each component brings essentially independent information to the discrimination method. Thus, using both together improves generalization ability of the combined classifier. Similar orthogonality holds for the other pairs of directions as well, suggesting that the different facial features tend to work independently in assisting with the overall good performance of DWD. Although none of the nose, lips, and chin give perfect separation, they all provide useful information to the overall discrimination. This is consistent with the much larger gap between the combined DWD subpopulations in Movie \hyperref[mov:projns:DWD]{2a}, relative to the gaps shown in Figure \ref{fig:ScoresMatrix}. Overall, we now have several new DWD directions each of which discriminate some part of the population well, but not everyone. In particular, some women fall into that category because of their smile, others because of their eyebrows.

\subsection{Classification errors for DWD, SVM, and FLD}
\label{sec:exp:3}

The last experiment evaluates the performance of DWD, SVM, and FLD (with Moore-Penrose pseudo-inverse for the covariance matrix) in the classification of male and female faces. To that aim, we classify an independent testing dataset with the above classifiers, after training them with the above studied $n=108$ images. The testing dataset consists of $100$ images, $58$ men and $42$ women, which were obtained under similar illumination conditions and viewpoint as the training data. Similarly as for the training dataset, there was a pre-processing step for registering the images using a generalized Procrustes analysis.

\begin{figure}[!htpb]
\centering
\includegraphics[width=0.85\textwidth]{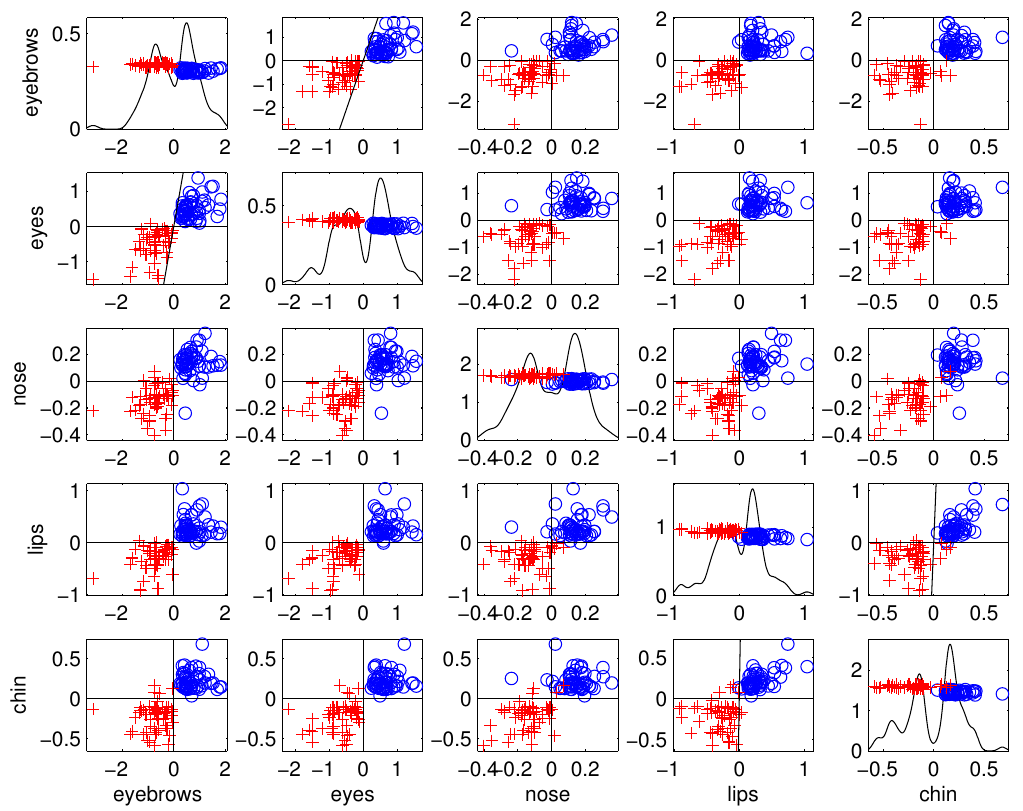}
\caption{\small Projection of the full image vectors
onto the DWD directions defined by the most important regions. The
plots along the diagonal show the one-dimensional projection onto
each direction. The off-diagonal plots are two-dimensional projections
by pairs. The symbols and colors are as in Movie \ref{mov:projns}, and black lines represent the two DWD directions. It shows that eyebrows and eyes give the most male-female discrimination, and others contribute weaker, but nearly independent additional information.\label{fig:ScoresMatrix}}
\end{figure}

\begin{table}[!htpb]
\centering%
\begin{tabular}{|l|c|c|c|}
\toprule
\toprule
\textbf{Classification matrix} & \textbf{DWD}  & \textbf{SVM}  & \textbf{FLD} \\\midrule
Men classified as men  & $58$  & $57$  & $56$ \\\midrule
Women classified as women  & $40$  & $38$  & $15$ \\\midrule
Men classified as women  & $0$  & $1$  & $2$ \\\midrule
Women classified as men  & $2$  & $4$  & $27$ \\\midrule\midrule
\textbf{Error rate} & $2\%$  & $5\%$  & $29\%$ \\\bottomrule\bottomrule
\end{tabular}
\caption{\small Classification error on the testing dataset for DWD, SVM, and FLD. It shows superior performance of DWD, which is consistent
with each of the above results.\label{tab:error}}
\end{table}

Table \ref{tab:error} shows the classification error rates (defined as number misclassified, divided by the total) in the classification of the testing dataset when the full image vectors are considered. The results are quite consistent with what we shown in the above sections: SVM is much better than FLD, and DWD is substantially better than SVM.

\section{Conclusions}

We have illustrated the benefits of DWD over SVM, FLD, and PCA in a challenging HDLSS discrimination problem: the classification of male and female faces from a dataset of facial images. When performing classification, SVM exhibited signs of overfitting in terms of data piling, FLD was significantly worse than SVM, and DWD showed the best performance, both in terms of classification error and in terms of a wider, Gaussian-shaped and interpretable separation of subpopulations. PCA showed some discrimination ability, but it was clearly overcome by any of the above classifiers.\\

The classification by DWD automatically identified key features for male-female discrimination that contributed nearly orthogonal DWD directions to the classifier: eyebrows, eyes, nose, lips, chin, and the outline of the head. Indeed, each of the regions of pixels associated to eyebrows and eyes delivered perfect DWD classification in the training dataset. Careful interpretation of the DWD separating vector showed its success in capturing well-known physical differences between male and female faces, and also revealed other characteristic features of male and female facial images that were less immediate, such as the differences of smiles (women tend to smile more) and eyes (women's eyes are more open).

\section*{Acknowledgements}

J. S. Marron acknowledges support from the National Science Foundation under Grant No. 633074, E. Garc\'ia-Portugu\'es from project MTM2016-76969-P, and D. Pe\~{n}a from project ECO2015-66593-P, both last projects funded by the Spanish Ministry of Economy, Industry and Competitiveness, and ERDF funds.


\begin{thebibliography}{}

\bibitem[Ahn and Marron, 2010]{Ahn2010}
Ahn, J. and Marron, J.~S. (2010).
\newblock The maximal data piling direction for discrimination.
\newblock {\em Biometrika}, 97(1):254--259.

\bibitem[Belhumeur et~al., 1997]{Belhumeur1997}
Belhumeur, P.~N., Hespanha, J.~P., and Kriegman, D.~J. (1997).
\newblock Eigenfaces vs. {F}isherfaces: Recognition using class specific linear
  projection.
\newblock {\em IEEE Trans. Pattern Anal. Mach. Intell.}, 19(7):711--720.

\bibitem[Benito et~al., 2004]{Benito2004}
Benito, M., Parker, J., Du, Q., Wu, J., Xiang, D., Perou, C.~M., and Marron,
  J.~S. (2004).
\newblock Adjustment of systematic microarray data biases.
\newblock {\em Bioinformatics}, 20(1):105--114.

\bibitem[Benito and Pe\~{n}a, 2005]{Benito2005}
Benito, M. and Pe\~{n}a, D. (2005).
\newblock A fast approach for dimensionality reduction with image data.
\newblock {\em Pattern Recognit.}, 38(12):2400--2408.

\bibitem[Bruce, 1988]{Bruce1988}
Bruce, V. (1988).
\newblock {\em Recognising Faces}.
\newblock Essays in cognitive psychology. Lawrence Erlbaum Associates, Inc.,
  Hillsdale, NJ.

\bibitem[Bruce et~al., 1992]{Bruce1992}
Bruce, V., Burton, A.~M., and Craw, I. (1992).
\newblock Modelling face recognition.
\newblock {\em Phil. Trans. R. Soc. Lond. B}, 335(1273):121--127.

\bibitem[Burton et~al., 1993]{Burton1993}
Burton, A.~M., Bruce, V., and Dench, N. (1993).
\newblock What's the difference between men and women? {E}vidence from facial
  measurement.
\newblock {\em Perception}, 22(2):153--176.

\bibitem[Carmichael and Marron, 2017]{Carmichael2017}
Carmichael, I. and Marron, J. (2017).
\newblock Geometric insights into support vector machine behavior using the
  {KKT} conditions.
\newblock {\em arXiv:1704.00767}.

\bibitem[Farkas et~al., 1987]{Farkas1987}
Farkas, L.~G., Munro, I.~R., and Kolar, J. (1987).
\newblock Relationships of profile segment inclinations in the faces of north
  american caucasians.
\newblock In Farkas, L.~G. and Munro, I.~R., editors, {\em Anthropometric
  Facial Proportions in Medicine}, pages 67--78, Springfield, IL. Charles C.
  Thomas.

\bibitem[Galton, 1910]{Galton1910}
Galton, F. (1910).
\newblock Numeralized profiles for classification and recognition.
\newblock {\em Nature}, 83:127--130.

\bibitem[Genton et~al., 2015]{Genton2015}
Genton, M.~G., Castruccio, S., Crippa, P., Dutta, S., Huser, R., Sun, Y., and
  Vettori, S. (2015).
\newblock Visuanimation in statistics.
\newblock {\em Stat}, 4(1):81--96.

\bibitem[Gizatdinova and Surakka, 2010]{Gizatdinova2010}
Gizatdinova, Y. and Surakka, V. (2010).
\newblock Automatic edge-based localization of facial features from images with
  complex facial expressions.
\newblock {\em Pattern Recognit. Lett.}, 31(15):2436--2446.

\bibitem[Gunn, 1998]{Gunn1998}
Gunn, S.~R. (1998).
\newblock Support vector machines for classification and regression.
\newblock {\em Department of Electrical and Computer Science, University of
  Southampton}.

\bibitem[Hall et~al., 2005]{Hall2005}
Hall, P., Marron, J.~S., and Neeman, A. (2005).
\newblock Geometric representation of high dimension, low sample size data.
\newblock {\em J. Roy. Statist. Soc. Ser. B}, 67(3):427--444.

\bibitem[Hastie et~al., 2009]{Hastie2009}
Hastie, T., Tibshirani, R., and Friedman, J. (2009).
\newblock {\em The Elements of Statistical Learning}.
\newblock Springer Series in Statistics. Springer, New York, second edition.

\bibitem[Kawulok et~al., 2016]{Kawulok2016}
Kawulok, M., Celebi, M.~E., and Smolka, B. (2016).
\newblock {\em Advances in Face Detection and Facial Image Analysis}.
\newblock Springer Publishing Company, Switzerland.

\bibitem[Marron, 2015]{Marron2015}
Marron, J.~S. (2015).
\newblock Distance-weighted discrimination.
\newblock {\em Wiley Interdiscip. Rev. Comput. Stat.}, 7(2):109--114.

\bibitem[Marron et~al., 2007]{Marron2007}
Marron, J.~S., Todd, M.~J., and Ahn, J. (2007).
\newblock Distance-weighted discrimination.
\newblock {\em J. Am. Stat. Assoc.}, 102(480):1267--1271.

\bibitem[Pe\~na and Prieto, 2000]{Pena2000}
Pe\~na, D. and Prieto, F.~J. (2000).
\newblock The kurtosis coefficient and the linear discriminant function.
\newblock {\em Stat. Probab. Lett.}, 49(3):257--261.

\bibitem[Pe\~na and Prieto, 2001]{Pena2001}
Pe\~na, D. and Prieto, F.~J. (2001).
\newblock Cluster identification using projections.
\newblock {\em J. Am. Stat. Assoc.}, 96(456):1433--1445.

\bibitem[Samal and Iyengar, 1992]{Samal1992}
Samal, A. and Iyengar, P.~A. (1992).
\newblock Automatic recognition and analysis of human faces and facial
  expressions: A survey.
\newblock {\em Pattern Recognit.}, 25(1):65--77.

\bibitem[Turk and Pentland, 1991]{Turk1991}
Turk, M. and Pentland, A. (1991).
\newblock Eigenfaces for recognition.
\newblock {\em J. Cognit. Neurosci.}, 3(1):71--86.

\bibitem[Valentin et~al., 1994]{Valentin1994}
Valentin, D., Abdi, H., O'Toole, A.~J., and Cottrell, G.~W. (1994).
\newblock Connectionist models of face processing: A survey.
\newblock {\em Pattern Recognit.}, 27(9):1209--1230.

\bibitem[Vapnik, 1995]{Vapnik1995}
Vapnik, V.~N. (1995).
\newblock {\em The Nature of Statistical Learning Theory}.
\newblock Springer-Verlag, New York.

\bibitem[Wu and Huang, 1990]{Wu1990}
Wu, C.~J. and Huang, J.~S. (1990).
\newblock Human face profile recognition by computer.
\newblock {\em Pattern Recognit.}, 23(3):255--259.

\bibitem[Yang and Yang, 2003]{Yang2003}
Yang, J. and Yang, J. (2003).
\newblock Why can {LDA} be performed in {PCA} transformed space?
\newblock {\em Pattern Recognit.}, 36(2):563--566.

\bibitem[Zhao et~al., 2003]{Zhao2003}
Zhao, W., Chellappa, R., Phillips, P.~J., and Rosenfeld, A. (2003).
\newblock Face recognition: A literature survey.
\newblock {\em ACM Comput. Surv.}, 35(4):399--458.

\end{thebibliography}

\end{document}